\documentclass[twoside,twocolumn]{article}
\usepackage{color}
\usepackage{CJK}
\usepackage{longtable,booktabs}
\usepackage{ltablex}
\usepackage{graphicx}
\usepackage{amsmath}
\usepackage{amssymb}
\usepackage{tabularx}
\usepackage{url}  
\usepackage[T1]{fontenc}                    
\usepackage[bitstream-charter]{mathdesign}
\usepackage[english]{babel}                 

\usepackage[a4paper,left=1.5cm,right=1.5cm,top=2.0cm,bottom=2.5cm]{geometry} 
\usepackage{lettrine}                       

\usepackage[hang, small, labelfont=bf, up, textfont=it, up]{caption}  
\usepackage{booktabs}                       

\usepackage{titling}                        
\usepackage[noblocks]{authblk}
\usepackage{hyperref}                       
\usepackage{xcolor}

\usepackage{abstract}                       

\newcommand{\upcite}[1]{\textsuperscript{\textsuperscript{\cite{#1}}}}

\newcommand\ion[2]{#1$\;${\scriptsize\rmfamily\uppercase\expandafter{\romannumeral#2}}\relax}

\pretitle{\begin{center}\Huge} 
\posttitle{\end{center}} 

\title{\textbf{V-Dynamics: True Threshold for Infectious Disease Dynamics
 and Synchronizing Human Defenses with Virus Evolution
}}
\author[ ]{\small Yong-Shou Chen}
\affil[ ]{\small China Institute of Atomic Energy, Beijing 102413, China}

\date{} 

\begin{document}

\maketitle

\small

\noindent\textbf{The COVID-19 pandemic exposed critical gaps in predicting viral emergence and trajectory. Although the century-old $R_0$ has been epidemiology's central metric, its limitations as an average for heterogeneous dynamic systems are widely accepted, and inconsistencies between the $R_0=1$ threshold and data are documented. This motivates the personal basic reproductive number $R_P=exp(\chi\eta_0d)-1$, defined in V-Dynamics as the true threshold for infectious disease. $R_P=0$ is a pathogen-intrinsic, pre-outbreak threshold replacing the population-averaged $R_0=1$ with a physically grounded criterion for outbreak onset. $R_P$ explains key epidemic phenomena $R_0$ cannot, including superspreading and multi-wave events, and identifies superspreaders before transmission occurs. V-Dynamics, a physics-based framework, models host-virus interaction as a synchronized dual chaotic system. At its core, the viral exponent $\eta$, derived from viral replication, immune neutralization, and drug-interaction cross sections, governs the entire infectious process, dictating both intra-host viral load phases-peak, plateau, clearance, and chaotic fatal runaway-and, through a scaling law, the Lyapunov exponent governing population-level transmission dynamics. Retrospective validation with SARS-CoV, SARS-CoV-2, and Omicron outbreaks, and with RSV vaccine efficacy, confirms the model's predictive capacity. We further show that $R_P$, as a practical point-of-care triage tool, identifies superspreaders before they transmit, using longitudinal SARS-CoV-2 viral load data. Crucially, V-Dynamics enables pre-outbreak prediction via in vitro measurement of viral reaction cross sections, offering a pathway to proactive pandemic defense. By integrating quantum-mechanical interaction models with chaos theory across scales, this framework provides a quantitative roadmap for anticipating and preempting future viral threats.}

\vspace{10pt}
\noindent The persistent threat of viral pandemics-exemplified by the devastating global impact of COVID-19-highlights a fundamental scientific and public health challenge: our inability to predict the emergence, transmission, and virulence of novel viruses before they spread widely. Despite advances in genomic sequencing, molecular virology, and population-level analyses based on the basic reproductive number $R_0$, the current paradigm remains largely reactive and inadequate for forecasting viral behavior prior to widespread outbreaks \upcite{Kimura2026, Junes2024}.
Although $R_0$ has served as epidemiology's central metric since the SIR model of epidemic spread a century ago \upcite{Kermack1927}, the criticism that it is an average quantity unsuitable for heterogeneous dynamic systems is widely accepted, and the threshold $R_0=1$ has well-documented inconsistencies between model and data \upcite{Breban2007}. Central to our current understanding of epidemic spread is the basic reproductive number, $R_0$, defined as the average number of secondary infections caused by a single infected individual in a fully susceptible population. However, $R_0$-based approaches oversimplify the complex heterogeneous dynamics governing viral spread and cannot accurately forecast outbreak trajectories, as identical $R_0$ values can yield vastly different epidemic outcomes \upcite{Delamater2021}. These inherent limitations were starkly revealed during the SARS-CoV-2 pandemic, when high transmissibility and immune evasion emerged unexpectedly, leading to unprecedented morbidity, mortality, and socioeconomic disruption.
One long-standing and striking feature in the study of epidemics is the absence of a true first-principles threshold and, consequently, the lack of prediction and precise control of outbreak emergence. This is precisely the fundamental limitation of traditional $R_0$ models (since 1927) and their modern extensions, as starkly exposed during the COVID-19 pandemic when their "well-mixed" homogeneity assumption broke down. A question arises that classical theory cannot answer: what is the true threshold for an infectious disease? We argue that the answer lies not in the homogeneous mixing assumption but in the stability boundary of the viral replication system.
To address these limitations and move from descriptive to predictive science, we introduce V-Dynamics, a novel interdisciplinary framework that conceptualizes host-virus interactions as a synchronized dual chaotic system. This framework provides a mechanistic, physics-based foundation capable of integrating viral replication dynamics, host immune responses, and population-scale transmission into a unified model. We demonstrate that the viral exponent $\eta$, the viral growth rate, governs not only intra-host infection phases-peak ($\eta>0$), plate ($\eta\approx0$), and clearance ($\eta<0$)-as well as the viral production chaotic state (zero-shedding), but also, via a scaling law, determines the Lyapunov exponent of population-level transmission chaos. This establishes a predictive link between microscopic viral replication efficiency and macroscopic pandemic trajectory.
At the core of V-Dynamics, we introduce the personal basic reproductive number, $R_P=exp(\chi\eta_0d)-1$, as the true threshold parameter for infectious disease dynamics, derived from the biophysics of virus-host interaction. Unlike the classical $R_0$, which is a population-average statistic based on the homogeneous mixing assumption, $R_P$ is a measurable, mechanistic quantity, derived as the number of secondary infections caused by a specific infected individual in a heterogeneous population. $R_P>0$ predicts an outbreak, $R_P<0$ predicts decline, and $R_P=0$ corresponds to zero net growth, with $\eta_0=0$, $d=0$, or $\chi=0$ each defining a threshold. Because $R_P=0$ unites cell-level ($\eta_0=0$), individual-level ($d=0$), and population-level ($\chi=0$) thresholds, it explains key epidemic phenomena that $R_0$ cannot, including superspreading and multi-wave events. In particular, $R_P$ identifies superspreaders before transmission occurs, as slow cleaners have large $d$ and exponentially large $R_P$.
The century-old $R_0=1$ criterion is an emergent population average with no direct biophysical meaning, motivating a host-level physical threshold derived from the Lyapunov exponent of viral replication.
Superspreading is a well-documented feature of SARS-CoV-2: roughly 20\% of cases account for 80\% of transmission, and this over-dispersed distribution is quantified by the dispersion parameter $k$, where values below unity signal substantial superspreading potential \upcite{Smith2005, Wegehaupt2023}; behavioral and environmental explanations for transmission heterogeneity also abound. However, the extent to which heterogeneity in infectiousness arises from individual-level viral dynamics remains unknown. To address this gap, $R_P$ provides an individual-level explanation: the underlying features of virus shedding dynamics within individual hosts contribute essentially to superspreading phenomena. Critically, $R_P$ describes quantitatively the superspreading phenomenon as heterogeneity in inter-individual transmission potential and explains why and predicts when it occurs. To validate this transformative approach for identifying superspreaders long before they transmit virus, we calculate $R_P$ from daily longitudinal nasal sampling data of SARS-CoV-2 Omicron viral loads for 14 participants in a case-ascertained household transmission study in Los Angeles County \upcite{Winnett2023}. The results show great individual heterogeneity in transmission, and person D is identified as the most probable potential superspreader, having a uniquely large $R_P$.
Here, we present the theoretical foundation of V-Dynamics, derive its governing equations, and validate its predictive power through retrospective analysis of SARS-CoV, SARS-CoV-2, and Omicron variant outbreaks and of RSV vaccine efficacy, demonstrating the approach for identifying superspreaders before they transmit virus. We further propose a pathway toward pre-outbreak prediction and synchronizing human defenses with viral evolution via in vitro measurement of $\eta$, offering a transformative shift from descriptive to predictive virology. By bridging quantum mechanics, kinetic theory, and nonlinear dynamics, this framework provides a quantitative, actionable roadmap for anticipating and mitigating future viral threats.


\vspace{10pt}
\noindent\textbf{Central Concepts}

\noindent The virus(V)-dynamics is introduced from a physics perspective to explain virus (e.g. SARS-CoV-2) infection in human lungs and its pandemic spread. The classical SIR model of epidemiology is grounded in the chemical mechanism of the law of mass action, and used to describe macroscopic phenomena in social population. V-Dynamics, by contrast, rests on a fundamentally different philosophical and mathematical basis: it is a mechanistic, physics-driven, multi-scale framework formulated as an analytically derived dynamical system.

\noindent\textbf{Virion replication reaction.}
A virion replication reaction is a process in which a host cell (H-particle) is bombarded by a virion, undergoes transformation into a ruptured H-particle with progeny virions.
Virus strikes and attaches to a host cell, passes through the cell membrane, forming a composite cellular system. Within this composite, new virions are produced and subsequently released.

\noindent\textbf{Self-sustaining virion chain reaction.}
This is a viral replication process in which virions released from one H-particle infect and destroy other H-particles, releasing more virions in a continuous, self-amplifying cascade. The reaction becomes self-sustaining when, on average, each replication event produces enough virions to trigger at least one more replication event. This chain reaction is often controlled through immune responses, where antibodies (A-particles) neutralize virions, establishing a transient balance between virion production and destruction. Experimental studies using inoculated human lung tissue explants with SARS-CoV-2 or SARS-CoV have observed such self-sustaining chain reactions, with virion counts increasing a hundredfold within 2 to 48 hours for SARS-CoV-2 \upcite{Chu2020}.

\noindent\textbf{Virions as Brownian particles.}
By examining viruses through a physical lens, we treat virions as Brownian particles undergoing random thermal diffusion in the viscous medium (bodily fluids) surrounding cells. Virus-cell interactions are quantified using reaction cross sections, rooted in quantum theory, that measure the rate at which a given type of interaction event occurs.
As a result, while traditional virology can describe viral phenomena with precision, it lacks predictive capability. In contrast, V-Dynamics enables quantitative prediction of the rates at which these phenomena occur.
%

\vspace{10pt}
\noindent\textbf{General Theory}

\noindent Based on the concepts of V-Dynamics we derive central equations, including the viral kinetic equation, burnup equation of H-particle, equation of viral production chaos, and equation of population transmission chaos (see Methods). We define the epidemic threshold parameter, $\eta$, governing the entire infectious disease process, and quantifiable through microscopy and in vitro experiments. We further derive the personal basic reproductive number, $R_P$, as the true threshold for infectious disease dynamics, which unites cell-level, individual-level, and population-level thresholds.(See Methods for all details)

\noindent\textbf{Viral kinetic equation.} The physical picture is to treat virions as Brownian particles undergoing random thermal motion in viscous medium (bodily fluids) surrounding the cell, and interaction with the cells. Starting from the Brownian diffusion equation we derive the viral kinetic equation:
\begin{equation}
  \label{kieq}
     \frac{dn_V(t)}{dt}=\eta n_V(t).
  \end{equation}
Where $n_V(t)$ is the number density of virions, or the viral load, $\eta=\rho/
\ell$ is the  viral exponent. The reactivity $\rho$ and the generation time $\ell$ are expressed as:
\begin{equation}
  \label{kipa}
     \rho=\frac{k_{eff}-1}{k_{eff}}, \ell=\frac{\ell_{0}}{k_{eff}}.
  \end{equation}
Where $k_{eff}$ is the effective multiplication constant (ratio of virions produced to those lost), and $\ell_0$ is the virion effective lifetime.
For an initial condition $n_V^0=n_V(t=0)$ and time-dependent $\eta(t)$, the solution is:
 \begin{equation}
  \label{so1}
     n_V(t)=n_V^{0}exp\left(\int_{0}^{t}\eta(t)dt\right).
  \end{equation}
We define the viral exponent (VE), i.e., $\eta(t)=\rho(t)/\ell$, which specifies the exponential growth rate of virions.
For the basic viral exponent mode $\eta(t)=\beta-\gamma{t}$, the solution is analytical:
\begin{equation}
  \label{so3a}
     n_V(t)=n_V^{0}exp\left(\beta{t}-\frac{\gamma}{2}t^{2}\right).
  \end{equation}
We define $\eta(t)$ as a piecewise linear function over three distinct phases of infection:
\begin{equation}
  \label{etai}
  \eta_i(t)=\beta_i-\gamma_i{t}.
  \end{equation}
Where,
$i=1$: Peak/Exponential Growth Phase ($\gamma=0, \beta>0$);
$i=2$: Peak/Exponential Decay Phase ($\gamma>0, \beta<0$);
$i=3$: Plateau Phase ($\gamma\approx0, \beta\approx0$), viral load stays smooth);
$i=4$: Clearance/Decay Phase ($\gamma>0, \beta<0$), viral load declines).
The general solution for the viral load during any phase $i$, starting from initial viral load $n_{i-1}$ at time $t_{i-1}$, is:
\begin{equation}
  \label{so3b}
     n_V(t)=n_{i-1}exp\left(\beta_i(t-t_{i-1})-\frac{1}{2}\gamma_i(t-t_{i-1})^{2}\right).
  \end{equation}

\vspace{10pt}
\noindent\textbf{Viral Exponent $\eta$.}
\noindent The reactivity $\rho$ and generation time $\ell$, defined in Eq.(\ref{kipa}), are determined by the virion interaction cross sections,
$\rho=1-(\sigma^A_aN_A+\sigma^D_aN_D+DB^2)/\nu\sigma_rN_H$ and $\ell=1/\upsilon\nu\sigma_rN_H$, by substituting $\eta=\rho/\ell$, the viral exponent $\eta$ is expressed as:
\begin{equation}
\label{rhorate}
\eta=\upsilon(\nu\sigma_rN_H-\sigma_a^AN_A-\sigma_a^DN_D-DB^2),
\end{equation}
where $\upsilon$ is the thermal velocity of the virion, $D$ is the diffusion coefficient, and $\nu$ is the average number of virions produced per replication event. The quantities $\sigma_r$, $\sigma_a^A$, and $\sigma_a^D$ are the microscopic cross sections for replication, antibody neutralization, and drug absorption, respectively, while $N_H$, $N_A$, and $N_D$ denote the number densities of host cells, antibodies, and drugs. This formulation is notable for its physical transparency and structural elegance. Although the viral load dynamics is captured by the single scalar $\eta$, the underlying system is a high-dimensional nonlinear process involving viral mutation, host immune response, and drug intervention. The decomposition of $\eta$  into four competing mechanistic terms-each corresponding to a distinct microscopic cross section-effectively transforms the "black-box parameters" into interpretable, experimentally accessible components.
The proliferation term, $\nu\sigma_rN_H$, describes the sequence of virion-host cell binding, replication, and progeny release, and contributes positively to $\eta$, promoting outbreak. The antibody neutralization term, $\sigma_a^AN_A$, corresponds to antibody, or vaccine-mediated virion clearance, and contributes negatively, suppressing viral growth. The drug absorption term, $\sigma_a^DN_D$ , represents drug-induced virion inactivation, also contributing negatively. Finally, the loss term, $DB^2$, accounts for viral decay via leakage, providing an additional dissipative contribution.
This "all competition in one equation" structure bears a striking mathematical resemblance to the criticality condition in neutron transport theory of reactor dynamics, where $k_{eff}=1$. Analogously, viral exponential growth is governed by $\eta$, when $\eta=0$ (or $k_{eff}$=1) the system resides at a critical state-a condition structurally identical to the criticality concept in reactor physics.
Viral infection is a complex nonlinear process in which viral replication and shedding dynamics are shaped by the complex interplay between host and viral factors, as implied in the $\eta$ equation. Crucially, the formulation offers experimental tractability. The cross-sections $\sigma_i$ can be determined through in vitro assays, which is the key to translating V-Dynamics from theoretical framework to engineering application.

\noindent\textbf{Burnup equation of H-particles.}
This equation describes the decline in H-particle concentration due to viral replication, with cellular burnout severity linked to clinical mortality. Neglecting feedback effects, it is written as:
\begin{equation}
  \label{burnu2}
     \frac{dN_{H}(t)}{dt}=-\sigma_{a}\phi N_{H}(t),
\end{equation}
where $N_H$ is H-particle density, $\sigma_a=\sigma_r+\sigma_c$ is the virion absorption cross section, and $\phi=\upsilon n_V$ is the virion flux in the lung core. Fuel consumption is defined as $N_{H}/N_{H}^{0}$, and $N_{H}^{0}=N_H(t=0)$.

\noindent\textbf{Equation of viral production chaos.}
The chaotic dynamics of viral production, represented by the viral load $n_{V}(t)$, is modeled by
\begin{equation}
\label{npos2}
 n_{V}(h)=\frac{n_0 n_\infty}{n_0+n_\infty exp(-\lambda_V{h})}.
\end{equation}
Where the Lyapunov Exponent (LE) equals the initial viral exponent, i.e., $\lambda_V=\eta_0$, $h$ (in hours) represents the time from
the start of the infection, $n_0=n_{V}(t=0)$, and $n_\infty=n_{V}(t\rightarrow\infty)$ is a finite saturation value.

\noindent\textbf{Equation of population transmission chaos.}
Integrating viral dynamics with chaos theory yields a population-level transmission equation for the number of positive cases $N_{pos}$:
\begin{equation}
\label{npos1}
 N_{pos}(d)=\frac{N_0 N_\infty}{N_0+N_\infty exp(-\lambda{d})}.
\end{equation}
Where $\lambda$ is the Lyapunov Exponent, $d$ (in days) represents the time from
the start of the infection spread, $N_0=N_{pos}(t=0)$, and $N_\infty=N_{pos}(t\rightarrow\infty)$ is a finite saturation value.
The two LEs satisfy the scaling law: $\lambda=\chi\lambda_V=\chi\eta_0$, where $\chi$ is the empirically calibrated scaling constant for the specific virus and social setting (e.g., $\chi\sim0.03$ for SARS-CoV-2 pre-vaccination).

\noindent\textbf{Personal basic reproductive number.}
The personal basic reproductive number, $R_P$, is defined as the number of secondary infections generated by a specific infected individual over their infectious period. It is derived from the population transmission equation, Eq.(\ref{npos1}), as:
$R_P=\lim_{N_\infty\rightarrow\infty}N_\infty/(1+N_\infty exp(-\lambda{d}))-1=e^{\lambda{d}}-1$,
where $d$ (in days) is the specific infectious period. Substituting $\lambda=\chi\eta_0$ yields,
\begin{equation}
\label{r0p}
 R_P=e^{\chi\eta_0{d}}-1.
\end{equation}
The subtractive term -1 removes the index case $N_0=1$.
In traditional epidemiology, $R_0$ is an average quantity, the mean number of secondary cases produced by a typical infected individual in a fully susceptible population. While useful for population-level planning, this average obscures substantial individual-level heterogeneity in infectiousness. By contrast, in the V-Dynamics framework, transmission follows a deterministic chaotic process (logistic-like growth), such that the secondary cases arising from a single index case are not a linear average but rather the integrated outcome of a chaotic trajectory over the specific infectious period $d$.
In Eq.(\ref{r0p}), $\chi$ is an empirically calibrated constant that encapsulates all population-level extrinsic factors (e.g., contact patterns, environmental conditions, intervention measures). Consequently, $R_P$  becomes a pure function of the patient's own viral kinetics. Crucially, both $\eta_0$ (the initial Lyapunov exponent, representing the true epidemic threshold) and $d$ are extracted directly from the patient's viral load trajectory $n_V(t)$. Thus, $R_P$ reflects an intrinsic property of that specific host-pathogen interaction. Clinically, $\eta_0$  is obtained from the log-linear slope of the patient's early viral load-values, and $d$ from the duration during which viral load remains above the infectious threshold. This enables $R_P$ to serve as a personalized biophysical barcode of transmission capacity. By translating theoretical chaos dynamics into a real-time, patient-specific metric, it offers a practical triage tool for identifying potential superspreaders at the point of care, well before they have the opportunity to infect others.

\noindent\textbf{True threshold $R_P$ in epidemiology.}
We introduce the first-principles epidemic threshold that identify superspreaders before they transmit.
Since the SIR model a century ago \upcite{Kermack1927}, the basic reproductive number $R_0$-the mean secondary infections from a single case in a fully susceptible population-has been epidemiology's central metric. However, the criticism that $R_0$ is an average quantity, unsuitable for heterogeneous dynamic systems, is widely accepted, and the threshold $R_0=1$ has well-documented inconsistencies between model and data \upcite{Breban2007}. This limitation was starkly exposed during COVID-19, when the homogeneity assumption broke down.
V-Dynamics introduces instead the personal basic reproductive number $R_P$, which is pathogen-intrinsic, pre-outbreak, and mechanistically grounded. Its core idea is to characterize transmission heterogeneity using a constant rooted in biophysical mechanisms, such as the dynamic balance between viral mutation and immune response. $R_P$, Eq.~\ref{r0p}, with a threshold at 0, is a true epidemic threshold because it is derived from the Lyapunov exponent ($\eta_0=0$), the fundamental stability boundary of the viral replication system. It shows that even when the classical $R_0$  is below 1, an epidemic may persist if individual transmission capacities within the transmission chain are correlated. Traditional $R_0=1$ is an emergent threshold arising from population-level averaging and has no direct biophysical meaning. This is why $R_P$ can identify superspreaders before they transmit, whereas traditional epidemiology can only identify them after an outbreak.
The two thresholds-$R_0=1$ and $R_P=0$-are not contradictory but reflect different levels of biological organization. $R_0=1$ is the population-level statistical threshold, an average over the entire population; its limitation is retrospective, since it can only be calculated after an outbreak and cannot identify who will transmit. $R_P=0$ is the host-level physical threshold, a property of the host-pathogen interaction; its advantage is prospective, since it can be calculated at diagnosis from early viral load data and can identify potential superspreaders before they transmit.
The thresholds are complementary: averaging individuals over $R_P$ recovers $R_0$, $<R_P>=R_0$, where the average is population-weighted and the equality is a conceptual bridge under specified mixing assumptions. $R_P$ is an individual capacity quantity, whereas $R_0$ is a population mean effective quantity. In $R_P=exp(\chi\eta_0d)-1$, the scaling constant $\chi$ links cell-level viral dynamics to population-level transmission through $\lambda=\chi\eta_0$; the conversion from individual $R_P$ to population $R_0$ is performed by population-weighted averaging. Because behavioral and environmental factors are absorbed into $\chi$, and $\chi$ enters the exponent, heterogeneity in $\chi$ can substantially enlarge $R_P$. By Jensen's inequality, $<R_P>\geq R_P(<\chi>)$, so a simple arithmetic mean is inadequate; a weighted average-weighted by contact rate, susceptibility, or transmission potential-is required to recover $R_0$.
The individual threshold is the biological substrate of the population threshold: a population cannot have $R_0>1$ unless at least some individuals have $R_P>0$. $R_P=0$ asks whether a specific individual is biologically capable of transmitting the virus; $R_0=1$ asks whether the outbreak will grow in the population. In one sentence: traditional $R_0=1$ tells you whether the fire will spread through the forest; V-Dynamics $R_P=0$ tells you whether a specific tree is burning.
The unified threshold $R_P=0$ occurs when $\eta_0=0$, $d=0$, or $\chi=0$. $\eta_0=0$ is the core threshold value: zero net virion growth implies no second-generation virus production and thus no possibility of transmission. When $d=0$, the effective transmission time is zero: although the person is infected, their viral load remains below the critical threshold $n_{threshold}$ , so they lack sufficient capacity to trigger transmission. Because $n_{threshold}$ depends on viral survival, behavior, climate, environment, and population structure, regular seasonal outbreaks are observed; when $n_{threshold}$ decreases, $d$ increases and $R_P$ grows exponentially, leading to rapid outbreak escalation. $\chi=0$ is a population-level critical point: no new generation of infection is produced, i.e., zero net growth. Because speed is a population-level critical parameter, when $\chi=0$ no epidemic can take off regardless of $\eta_0$ and $d$ (provided they are finite), which explains why some sparsely populated areas remain free of outbreaks even during a global pandemic.
Finally, $R_P$ characterizes individual heterogeneity in viral shedding dynamics (see subsection about identify superspreders).

%
%

\vspace{10pt}
\noindent\textbf{Dual Chaotic Systems and the Rate Scaling Law}

\begin{figure}[t]
\includegraphics[angle=0, width=10cm]{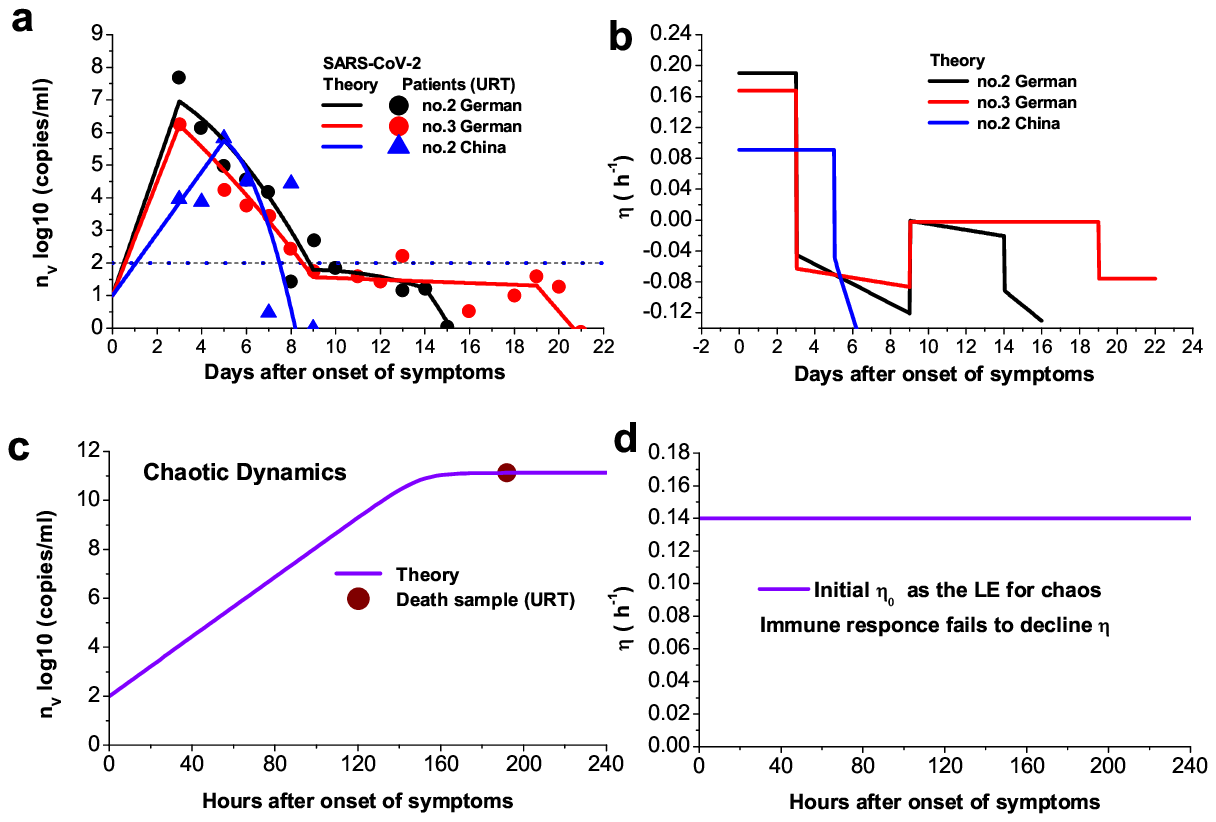} \\
\caption{Viral load dynamics of SARS-CoV-2 predicted by V-Dynamics.
\small \textbf{a} Theoretical viral load curves (solid lines) compared to URT data from patients in China (No. 2)  \upcite{Pan2020} and Germany (No. 2 and 3) \upcite{Wolfel2020}.  Dashed blue line indicates detection limit.
\textbf{b} Corresponding viral exponent $\eta$ (h$^{-1}$) driving the viral load phases: peak ($\eta>0$), plateau ($\eta\approx0$), and clearance ($\eta<0$).
\textbf{c} The chaotic viral load simulated using Eq.(\ref{npos2}) with $\lambda_V=0.14h^{-1}$, compared to upper respiratory tract (URT) data from patient who died \upcite{Pan2020}. \textbf{d} Corresponding viral exponent (persistent $\eta>0$) driving the chaotic dynamics.}
\label{sars2}
\end{figure}

\noindent The chaos theory proposes that diverse complex systems could share underlying commonalities, often describable through simple mathematical maps. A central concept in this framework is the Lyapunov Exponent (LE), which measures the rate at which initially proximate system trajectories converge (LE<0) or diverge (LE>0), quantifying the sensitivity of a dynamical system to small changes in state.
We propose that transmissible viruses operate within two intrinsically linked chaotic systems: the intra-host viral production system (within an infected individual) and the inter-host transmission system (within a population). We have formalized this dual-chaotic framework, deriving its governing equations (Eqs.\ref{npos2}, \ref{npos1}) and establishing a constitutive connection through an LE scaling law. In the absence of an effective immune response or clinical intervention, the intra-host system, driven by a persistently positive LE=$\eta_0$ (Fig.\ref{sars2}d), can enter a runaway chaotic state, culminating in an extremely high viral load saturation associated with patient death (Fig.\ref{sars2}c). For instance, a sputum sample from a fatal case on day 8 post-onset contained 1.34$\times10^{11}$ copies/ml \upcite{Pan2020}, whereas peak loads in non-fatal cases, typically observed around day 3 post-onset, are usually on the order of $\sim1\times10^{6}$ copies/mL (Fig.\ref{sars2}a).
In most non-fatal infections, viral loads deviate from pronounced chaotic growth, peaking around 3 days post-onset before declining sharply as the adaptive immune response activates, driving viral exponent $\eta$ from positive to negative (Fig.\ref{sars2}b). The critical window for preventing progression to a dangerous, high-chaos state is thus the first few days post-onset, prior to the system entering uncontrolled exponential growth. Optimized antiviral strategies can be designed to drive the intra-host Lyapunov Exponent $\lambda_V$ toward negative values, which would subsequently induce a synchronized reversal in the population transmission dynamics. The sign reversal of both Lyapunov Exponents would, within this framework, mark the end of a viral pandemic. V-Dynamics reveals that viral evolution adheres to a singular principle: maximizing the LE of production chaos by optimizing fitness ($\sigma_r$) while minimizing immune susceptibility ($\sigma_a^A$), thereby ensuring dominance over host defenses. Similarly, cross-species transmission is theorized to favor jumps between hosts with comparable LE values.

%
%

\vspace{10pt}
\noindent\textbf{Viral Replication and Shedding Dynamics}

\noindent Using the viral kinetic equation, we provide a quantitative description of SARS-CoV-2 dynamics by modeling virion interactions with H-particles (replication) and A-particles (immune response). Data from patients in China \upcite{Pan2020} and in Germany \upcite{Wolfel2020} were used for validation.

\noindent\textbf{Viral load and phases.}
The viral load trajectory is derived from Eq. (\ref{kieq}) through the viral exponent $\eta(t)$, whose values reflect the dynamic competition and balance between viral mutation and host immunity. This competition gives rise to a strongly nonlinear $\eta(t)$, characteristic of heterogeneous systems such as infectious disease epidemics. Specifically, we adopt a piecewise time-varying form for $\eta(t)$, where within each phase it follows the linear decay $\eta(t)=\beta-\gamma {t}$, which yields the viral load curve expressed in Eq. (\ref{so3b}). Our analysis resolves distinct phases of infection, driven by shifts in the host immune response. The calculated viral loads are validated against experimental data from the upper respiratory tract (URT), accurately capturing the characteristic peak, plateau, and clearance phases (Fig. \ref{sars2}a).
Importantly, the V-Dynamics framework identifies a high-risk regime-viral production chaos-which has been observed in a hospitalized COVID-19 patient. As shown in Fig. \ref{sars2}c, the chaotic phase is marked by an initial exponential growth that persistently approaches a saturation value, without the typical decline.
Phase transitions in this framework are governed by the switching of the viral exponent $\eta$. Specifically, the peak phase ($\eta>0$) corresponds to active viral growth, the plateau phase ($\eta\approx0$) represents a quasi-equilibrium state, and the clearance phase ($\eta<0$) indicates immune dominance. Shifts in $\eta$ are triggered by alterations in immune response or by pharmacological interventions.
Within the $\eta$ -theory, the onset of viral production chaos is determined by the sign of $\eta$.
In a typical, self-resolving infection, the immune system flips $\eta$ from positive to negative approximately 3 days post-infection. Thereafter in about 6 days, $\eta$ rebounds to near zero, before declining to negative values around 14 or 21 days post-infection (Fig. \ref{sars2}b). This sequence drives the viral load through the canonical three-phase trajectory: peak $\rightarrow$ plateau $\rightarrow$ clearance.
Conversely, if the immune response is delayed or ineffective, $\eta$ remains positive  (Figs.\ref{sars2}d). Under these conditions, the system enters a viral production chaotic attractor, characterized by the absence of peak, plateau, or clearance phases. Instead, the viral load continues to grow, approaching a saturation level, until the host's available target cells are exhausted (Figs. \ref{sars2}c).
This mechanistic framework explains why certain COVID-19 patients can reach viral loads as high as $10^{11}$ copies/mL in URT and eventually die. Crucially, this outcome is not stochastic, but rather a deterministic consequence of the system crossing a critical threshold into an irreversible chaotic branch. This insight provides a quantifiable theoretical basis for early risk stratification and defines a clear therapeutic window for antiviral interventions. This finding suggests that among COVID-19 deaths, some may have been caused by entering a chaotic runaway of viral production.

\begin{figure}[t]
\setlength{\abovecaptionskip}{-15pt}
\setlength{\belowcaptionskip}{-15pt}
\includegraphics[angle=0, width=15cm]{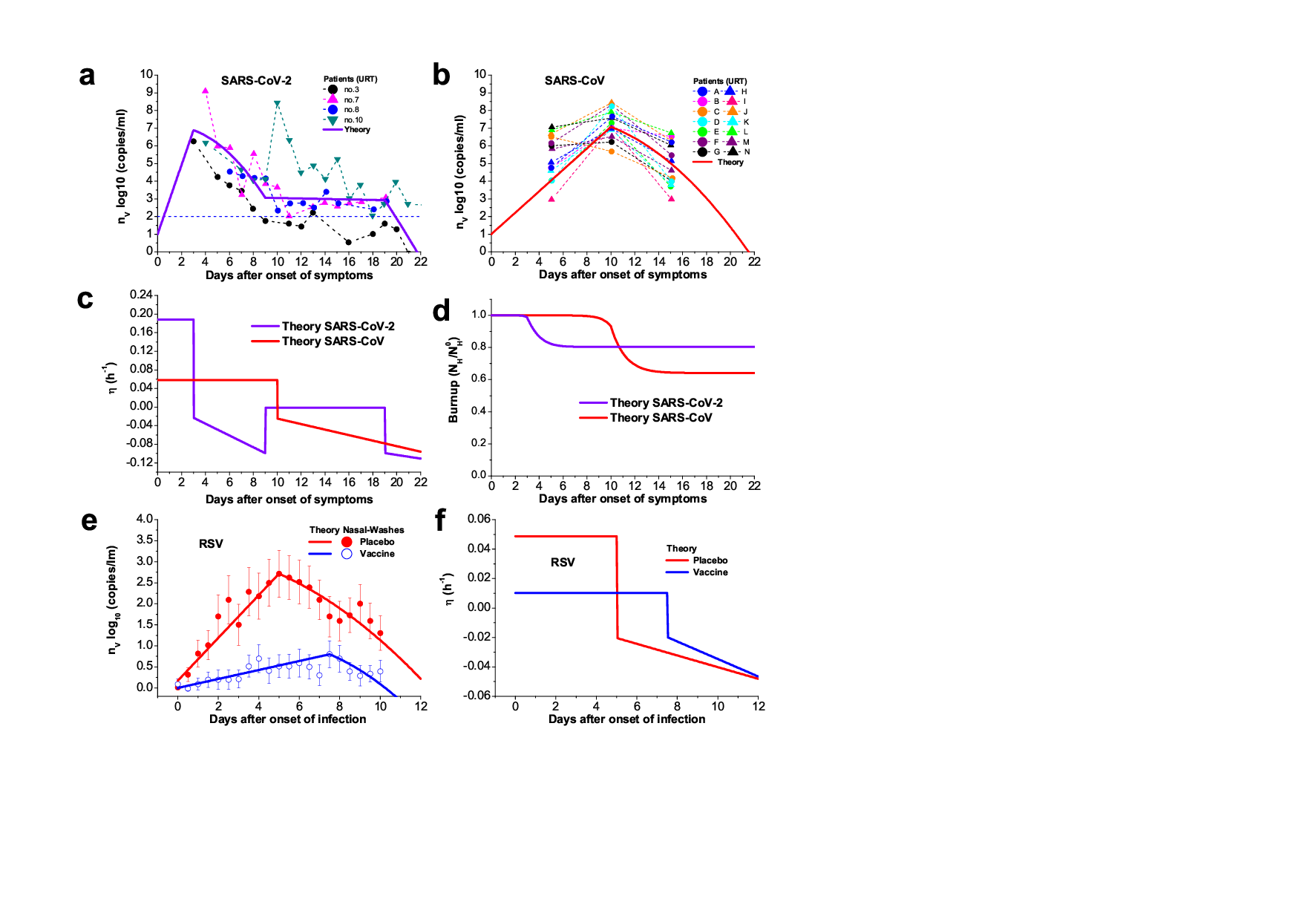}\\
\caption{Viral dynamics of SARS-CoV-2, SARS-CoV, RSV and Vaccine-RSV. \small \textbf{a} Modeled SARS-CoV-2 viral load (violet) versus URT data from four German patients  \upcite{Wolfel2020}. Detection limit shown as dashed line. \textbf{b} Modeled SARS-CoV viral load (red) versus data from 14 Chinese patients \upcite{Peiris2003}. \textbf{c} Viral exponent $\eta$ for two viruses, used to generate curves in \textbf{a} and \textbf{b}, respectively. SARS-CoV-2 exhibits a multi-step $\eta$ pattern and higher initial $\eta_0$, correlating with enhanced transmissibility. \textbf{d} Simulated host cell depletion. \textbf{e} Modeled viral load (curves) versus mean viral loads (from nasal washes) for the placebo group of 32 participants (solid circle) and for the MAV-BN-RS vaccine group of 31 participants (open circle) \upcite{Jordan2023}. \textbf{f} Corresponding viral exponents, $\eta$, used to generate the viral load curves in \textbf{e}.}
\label{sars-rsv}
\end{figure}

\noindent\textbf{Cell consumption (Burnup).}
Calculating H-particle burnup using Eq. (\ref{burnu2}) requires the viral load (virion flux) within the lung core. Direct measurement of this quantity remains challenging due to the invasive procedures involved. In contrast, upper respiratory tract (URT) viral loads represent time-accumulated quantities dependent on the duration of viable virus collection. It is therefore reasonable to assume that the lung core viral load is substantially lower than that of the URT yet follows a similar temporal pattern (Fig. \ref{sars-rsv} \textbf{a} and \textbf{b}).
For present purposes, a reduction factor for the core viral load may be freely chosen such that the calculated H-particle burnup for SARS-CoV-2 over the infection period is approximately 20 \%. Simulated lung-core viral loads (scaled down by 1/1000 from URT data) and a theoretical cross section of $\sigma_a=2.775\time10^{-11}cm^2$  were used to compute H-particle burnup (Fig. \ref{sars-rsv} \textbf{d}). SARS-CoV causes greater cellular damage, consistent with its higher lethality. Our calculations further indicate that the viral transmission rate of SARS-CoV-2 is approximately three times that of SARS-CoV, as the initial $\eta_0$ value of the former is roughly threefold that of the latter (Fig. \ref{sars-rsv} \textbf{c}), given that the Lyapunov exponent of population transmission chaos is proportional to $\eta_0$. These inter-viral differences stem from their distinct $\eta$ time courses (Fig. \ref{sars-rsv} \textbf{c}). Indeed, during the two outbreaks, SARS-CoV-2 has demonstrated higher transmissibility and a lower death rate than SARS-CoV \upcite{Johansson2021, Petersen2020}.
%

\noindent\textbf{Effects of vaccine.}
\noindent Respiratory syncytial virus (RSV) remains a common and recurring infectious disease. Infants are recognized as the highest-risk group, followed by older adults, who bear a substantial RSV disease burden \upcite{Nair2010}.
In a controlled challenge study, 31 participants received the MAV-BN-RS vaccine and 32 received a placebo. Mean viral loads measured from nasal washes were substantially lower in the vaccine group than in the placebo group \upcite{Jordan2023}. Here calculations accurately reproduce the viral load time-course data (Fig.\ref{sars-rsv}\textbf{e}), where the theoretical curves are generated using the corresponding viral exponents $\eta$ shown in Fig.\ref{sars-rsv}\textbf{f}. The V-Dynamics framework allows for a quantitative evaluation of vaccine efficacy through changes in the comparative $\eta$ values. The calculated results show that the initial $\eta_0$ is reduced by 80\% from the placebo to the vaccine group. Consequently, the Lyapunov exponents (LEs) for both viral production chaos and population transmission chaos could decline by approximately 80\%.  The $\eta$ sign shifts from positive to negative on day 5 post-infection for the placebo group while on day 7.5 for the vaccine group. The $\eta$ theory proves useful in quantitatively estimating the efficacy of vaccine as well as different antiviral therapies for virus infection.

\begin{figure}[t]
\setlength{\abovecaptionskip}{-15pt}
\setlength{\belowcaptionskip}{-15pt}
\includegraphics[angle=0, width=11cm]{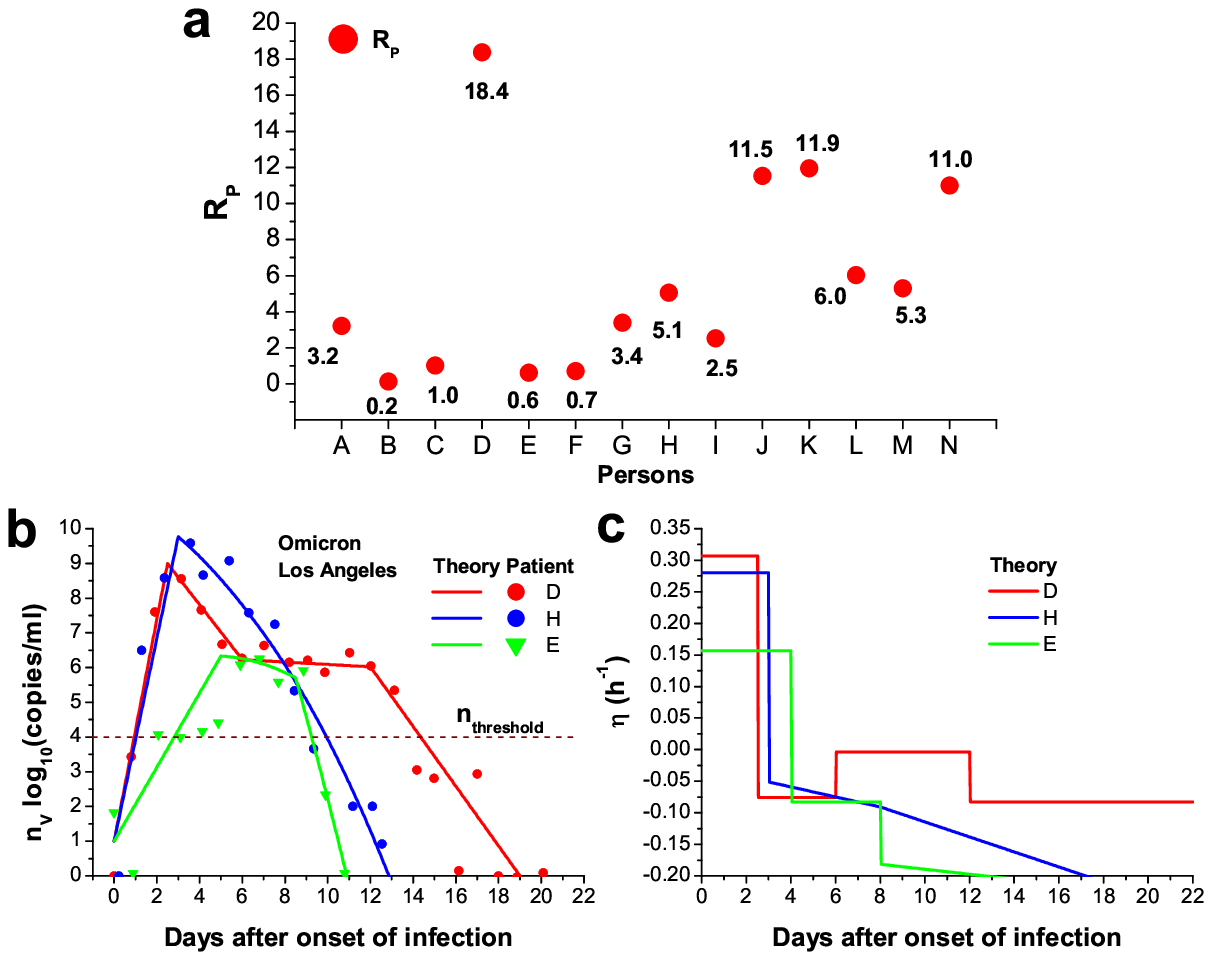}\\
\caption{Individual heterogeneity in transmission and identifying superspreaders. \textbf{a} The personal basic reproductive numbers $R_P$ for 14 individuals A-N (gender,age) in Los Angeles County, calculated from viral load timecourse data (nares swabs) of Omicron (acute SARS-CoV-2) infection \upcite{Winnett2023}, with $n_{threshold}=10^4 copies/lm$ and $\chi=0.03$. \textbf{b} Theoretical viral load curves (solid lines) compared to individual viral load data from participants D, H and E. \textbf{c} Corresponding viral exponents, $\eta(t)$, used to generate the viral load curves in \textbf{b}, indicating various shedding.}
\label{spr}
\end{figure}

\noindent\textbf{ $\eta_0$ as a metric characterizing pathogen variants.}
$\eta_0$ is essentially a normalized measure of pathogen virulence and of the largest Lyapunov exponent. Typical $\eta_0$ values may therefore serve as an intrinsic "epidemic strength" metric that scales with outbreak severity. For example, from Fig.\ref{sars-rsv}\textbf{c},\textbf{f} and Fig.\ref{spr}\textbf{c}, SARS-CoV-2 ($\eta_0=0.19 h^{-1}$) and Omicron ($\eta_0=0.23 h^{-1}$) correspond to extremely high epidemic strength (global pandemic); SARS-CoV ($\eta_0=0.06 h^{-1}$) to high epidemic strength (regional outbreak); RSV ($\eta_0=0.05 h^{-1}$) to moderate epidemic strength (seasonal pandemic); and vaccine-controlled RSV ($\eta_0=0.01 h^{-1}$) to low epidemic strength (under vaccine control).

%
%
\vspace{10pt}
\noindent\textbf{V-Dynamics Predictive Capacity and Accuracy}

\noindent V-Dynamics predicts virus transmission speed, pathogenicity, and impacts on public health, society, and economy. Its accuracy stems from modeling viral dynamics as a dual chaotic system, where Lyapunov Exponents are proportionally determined by the viral exponent $\eta$.

\noindent\textbf{Early-outbreak prediction.}
For a prediction at the early stage of outbreak, early viral load data are used to extract $\eta$, enabling estimation of transmission rate and virulence.
For example, retrospective simulations for SARS-CoV-2 and SARS-CoV show distinct $\eta$ patterns (Fig. \ref{sars-rsv} \textbf{c}), indicating significant different transmission and fatality rates, enabling real-time forecasting the emergence and risk of novel variants.
Yet, viral exponent $\eta$ monitoring enables phase-specific interventions, especially during the peak phase. We propose an AI-driven viral exponent Tracker (AI-VET) for real-time, personalized treatment adjustment.


\noindent\textbf{Pre-outbreak prediction.}
Pre-outbreak forecasting uses in vitro viral reaction cross section ($\sigma$) data to compute $\eta$, $\rho$, and $\ell$, enabling viral transmission, virulence, and cross-species transmission predictions, and forecasting pandemic trajectories. This represents a paradigm shift from descriptive to predictive science of virology.

%

\begin{figure}[t]
\setlength{\abovecaptionskip}{-15pt}
\setlength{\belowcaptionskip}{-15pt}
\includegraphics[angle=0, width=10cm]{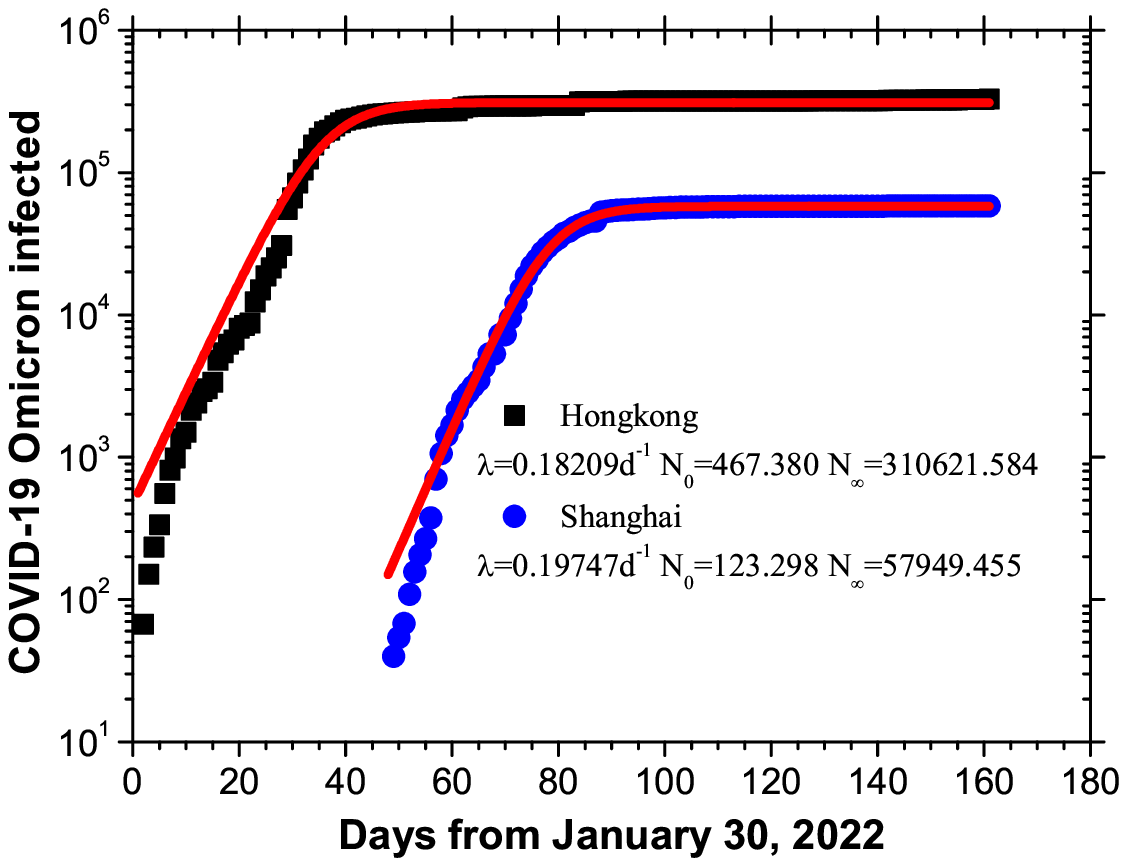}\\
\caption{Retrospective prediction of Omicron outbreaks in Hong Kong and Shanghai. Using Eq. (\ref{npos1}), modeled cumulative infections (curves) are compared to reported case data (points). Fitted saturation levels ($N_\infty$) reflect the differential impacts of intervention policies: strict lockdowns in Shanghai reduced $N_\infty$  by approximately $\sim81$ percent compared to Hong Kong, highlighting the trade-off between infection control and socioeconomic cost.}
\label{sprd}
\end{figure}

\noindent\textbf{Forecasting impacts of local outbreaks.}
Early viral exponent $\eta$ data enable simulation of transmission spread under different intervention policies, allowing prediction of the impacts of local outbreaks. To validate this approach, we retrospectively simulated the spread of SARS-CoV-2 Omicron and its negative impacts in Hong Kong and Shanghai. We fitted the parameters of Eq. (\ref{npos1}) to Omicron infection data from Hong Kong and Shanghai \upcite{HongShang2022}. Simulations of Omicron outbreaks in both cities (Fig. \ref{sprd}) show how intervention policies affect saturation infection levels, highlighting the trade-off between infection control and socioeconomic costs.

\vspace{10pt}
\noindent\textbf{Identify superspreaders long before they transmit virus}

\noindent Superspreading, where a small number of infected individuals cause a large share of transmission, is well documented for SARS-CoV-2, with roughly 20\% of cases responsible for 80\% of transmission, indicating an over-dispersed distribution of secondary cases \upcite{Smith2005,Ke2022}. This is quantified by the dispersion parameter $k$, as usual an additional parameter introduced in classical epidemiology, where values of ($k$ <1) indicate significant superspreading potential \upcite{Wegehaupt2023}.
While behavioral and environmental explanations for transmission heterogeneity abound, the extent of heterogeneity arising from individual-level viral dynamics remains unknown. To address this gap, the personal basic reproductive number, $R_P$, defined in V-Dynamics, quantitatively describes heterogeneity in infectiousness arising from individual-level viral dynamics. It offers an individual-level explanation: the underlying features of virus shedding dynamics within individual hosts contribute essentially to superspreading. Superspreading represents individual heterogeneity in infectious virus shedding driven by the timing and magnitude of the neutralizing antibody response. Critically, $R_P$ quantifies the superspreading phenomenon as heterogeneity in inter-individual transmission potential and explains why and predicts when it occurs.
The formulation $R_P=exp(\chi\eta_0d)-1$, Eq.\ref{r0p}, reveals a profound biological truth: a patient is not born a superspreader; they become one if $d$ is long. Even with moderate  $\eta_0$, if the immune system is slow to clear the virus (large $d$), $R_P$ becomes exponentially large. By monitoring $\eta_0$ and $d$ via serial testing, clinicians can identify "fast-rising and slow-clearing" patients who are as dangerous as high-viral-load patients and destined to become superspreaders. Extracting $\eta_0$ (from the log-linear slope of early viral load values) and $d$ (from the duration above the infectious threshold), $R_P$ serves as a personalized biophysical barcode of transmission capacity, a practical triage tool for identifying superspreaders at the point of care, long before they infect anyone.
To validate this idea, we calculate $R_P$ from daily longitudinal nasal sampling of SARS-CoV-2 Omicron viral loads for 14 participants (A-N) in a case-ascertained household transmission study in Los Angeles County \upcite{Winnett2023}.
For Three representative individuals, the viral load curves are calculated using Eq.\ref{so3b} and shown in Fig.\ref{spr}b, and the corresponding viral exponent $\eta(t)$ are shown in Fig.\ref{spr}c.
The infectious period $d$ is determined by solving $n_V (t)=n_{threshold}$ with $n_{threshold} =10^4 copies/mL$. Setting $\chi=0.03$, the calculated $R_P=18.4, 5.1$ and $0.6$ for D, H and E, respectively, as marked (Fig.\ref{spr}a). D is the most probable superspreader-with a distinctly large intrinsic transmission capacity, far larger $R_P$, and long $d$ (slow cleaner)-whereas E is a fast cleaner with short $d$. The $R_P$ values for all persons (Fig.\ref{spr}a) illustrate great individual heterogeneity in transmission. Person-to-person variation in infectious virus shedding, characterized by the $\eta$-pattern (Fig.\ref{spr}c), suggests that individual-level heterogeneity in viral and immune dynamics contributes to population-level transmission heterogeneity-superspreading.
In epidemic practice, transmission heterogeneity can be further amplified by heterogeneity in $\chi$ itself. Behavioral and environmental factors are absorbed into $\chi$; because $\chi$ enters the exponent, a modest increase in $\chi$ can substantially enlarge $R_P$. Since $R_P$ is convex in $\chi$, Jensen's inequality implies $<exp(\chi\eta_0d)>\geq exp(<\chi>\eta_0d)$, suggesting that heterogeneity in $\chi$ amplifies the population-level quantity relative to the mean-$\chi$ prediction. For example, potential superspreader D, with $R_P=18.4 (\chi=0.03)$, could become a superspreader with effective $R_P=51.0 (\chi=0.04)$ due to a high contact rate, consistent with behavioral and environmental explanations of superspreading, in which heterogeneity arises from interaction between highly infectious individuals and high-risk settings \upcite{Smith2005}.
A slow cleaner is a potential superspreader with large $d$ due to delayed immune response, whose $R_P$ becomes exponentially large. A zero cleaner is in a dangerous chaotic state of viral production, with the longest $d$ due to failed immunity, so $R_P$ is exponentially largest; as an ultra-superspreader, such a person will certainly die within weeks. A fast cleaner remains only in the peak phase, with relatively small $d$, and often recovers quickly; their transmission has a smaller $R_P$. An infinitely fast cleaner remains always before the peak phase, with zero $d$, innate immunity, and transmission capacity $R_P=0$.
%

%

\vspace{10pt}
\noindent\textbf{In Vitro Measurement of Virus-Cell Interactions}

\noindent V-Dynamics defines virus-cell interaction constants and proposes their quantification through in vitro experiments. Specifically, we aim to measure the cross sections for replication ($\sigma_r$), antibody neutralization ($\sigma_a^A$), and drug action ($\sigma_a^D$). Advanced biophysical techniques and AI-driven systems will be employed to observe single replication events, thereby establishing a foundation for predictive virology and antiviral design. Current in vitro measurements typically monitor viral load growth in inoculated tissues-using SARS-CoV-2 and SARS-CoV-capturing net replication dynamics without isolating the underlying interaction constants \upcite{Chu2020}. In their study, the authors reported that within a 48-hour interval, SARS-CoV-2 generated 3.20-fold more infectious virus particles than SARS-CoV from infected lung tissues. In fact, this experimental study has demonstrated the feasibility of the in vitro approach proposed by V-Dynamics for measuring virus-cell interactions.
For instance, by constructing a controlled microenvironment of host cells, virus, and antibodies/drugs within microfluidic channels, one can monitor viral load changes and back-calculate $\eta$. Moreover, by measuring individual cross-sections $\sigma_i$ and varying cell densities $N_i$, one can compute $\eta$  for specific populations or geographic regions. Once established, such in vitro platforms would enable rapid, a priori determination of $\eta$ for novel variants under laboratory biosafety conditions-precisely the technological foundation needed for pre-outbreak forecasting.


\vspace{10pt}
\noindent\textbf{Conclusion}

\noindent In summary, V-Dynamics establishes a physics-based, predictive framework for virology by modeling host-virus interaction as a dual chaotic system governed by the viral exponent $\eta$ and by the true epidemic threshold, the personal basic reproductive number $R_P$. We have shown that $\eta$ integrates microscopic viral replication, immune neutralization, and drug interaction into a single quantifiable metric that dictates both intra-host infection phases and population-level transmission dynamics. Through a scaling law linking viral production chaos ($\eta_0$) to transmission chaos ($\chi\eta_0$), together with $R_P$, V-Dynamics provides a mechanistic basis for forecasting pandemic trajectories from early viral load data or even from pre-outbreak in vitro measurements.
The individual threshold $R_P=0$ is the biological substrate upon which the population threshold $R_0=1$ is built: a population cannot have $R_0>1$ unless at least some individuals have $R_P>0$. Traditional $R_0=1$ tells you whether the fire will spread through the forest; V-Dynamics $R_P=0$ tells you whether a specific tree is burning. Because $R_P$ is measurable from early viral load time-course data, it converts the epidemic threshold from a retrospective population statistic into a prospective, individual-level diagnostic-one that can be calculated at the time of diagnosis and used to forecast outbreaks before they occur.
By deriving $R_P=exp(\chi\eta_0d)-1$ from individual viral load dynamics, this work shows that superspreaders are not born but made-identifiable at the point of care from the viral exponent $\eta_0$ and the infectious period $d$ long before they transmit-thereby converting superspreading from an unpredictable population-level phenomenon into a personalized, actionable triage tool for epidemic control. By extracting $\eta_0$ from the log-linear slope of early viral load values and $d$ from the duration above the infectious threshold, clinicians can identify "fast-rising and slow-clearing" individuals-who are just as dangerous as high-viral-load patients-and intervene before transmission occurs.
Moreover, because behavioral and environmental factors are absorbed into $\chi$ and  $\chi$ enters the exponent, heterogeneity in $\chi$ amplifies population-level transmission via Jensen's inequality, consistent with behavioral and environmental explanations of superspreading \upcite{Smith2005}. Together, these results transform theoretical chaos dynamics into a practical, life-saving triage tool: $R_P$ serves as a personalized biophysical barcode of transmission capacity, enabling superspreaders to be identified before they transmit a single virus. This approach has been validated by identifying superspreaders before they transmit, through application to longitudinal SARS-CoV-2 Omicron viral load data, which confirms pronounced individual heterogeneity, with slow cleaners exhibiting exponentially larger $R_P$ than fast cleaners. Retrospective validation with SARS-CoV, SARS-CoV-2, and Omicron outbreaks, and with RSV vaccine efficacy, confirms the model's capacity to differentiate transmission speed from lethality, predict wave progression, and evaluate intervention trade-offs. Crucially, the framework enables pre-outbreak risk assessment through experimental measurement of viral reaction cross sections, a move toward proactive rather than reactive pandemic preparedness. By tracking Lyapunov exponent (LE) values, this framework offers predictive insight into the directions of viral mutation and zoonotic transmission.
Looking forward, V-Dynamics opens several translational pathways: the development of an AI-driven viral exponent tracker (AI-VET) for real-time clinical monitoring and for identifying superspreaders before they transmit, the design of broad-spectrum antivirals based on optimized cross sections, and the integration of functional interaction data into global surveillance systems. While challenges remain in high-precision in vitro cross-section measurement through interdisciplinary collaboration, this paradigm shift toward predictive virology promises to synchronize human defense mechanisms with viral evolution. Ultimately, V-Dynamics redefines our approach to pandemic preparedness, offering a scientifically grounded, quantitatively predictive toolset for navigating future viral threats with greater foresight, precision, and resilience.

\vspace{10pt}
\noindent \textbf{Acknowledgements} {The author thanks Evelyn Jiang for valuable discussions on the SARS-CoV-2 pandemic. This work was supported by the Continuous Basic Scientific Research Project (WDJC-2019-13, BJ20002501) and the National Natural Science Foundation of China (11790325, 11975314).}

\noindent \textbf{Author contributions} {The author proposed, designed, and completed all aspects of this research.}

\noindent \textbf{Author Information} {The author declares no competing financial interests. Correspondence should be addressed to Y.-S. Chen (yshouchen@163.com).}

\appendix
\section*{Methods}

\small

\vspace{10pt}
\noindent\textbf{Virion Diffusion Equation}

\noindent In this physical picture, virions are treated as Brownian particles undergoing random thermal motion in the viscous medium (bodily fluids) surrounding cells, interacting with those cells. This behavior is described by the diffusion equation for virions. Starting from the Boltzmann transport equation and applying the isotropic and single-speed approximation, we derive the virion diffusion equation:
\begin{equation}
\label{dea}
  \frac{\partial n_V(\vec{r},t)}{\partial t}=D\upsilon\nabla^2n_V(\vec{r},t)-\Sigma_{a}\upsilon n_V(\vec{r},t)+S(\vec{r},t).
\end{equation}
Where $n_V(\vec{r},t)$ is the virion number density as a function of position $\vec{r}$ and time $t$, $D$ is the diffusion coefficient, $\upsilon$ is the average virion velocity, $\Sigma_a$ is the macroscopic absorption cross section, $S(\vec{r},t)$ is the source term accounting for virion production and external input.
Considering the production of progeny virions in replication, the source term can be expressed as:
  \begin{equation}
   \label{so}
    S(\vec{r},t)=k_\infty\Sigma_{a}\upsilon n_V(\vec{r},t)+S_0(\vec{r},t),
  \end{equation}
where the $k_\infty\Sigma_{a}\upsilon n_V(\vec{r},t)$ is the virion production rate, and $k_\infty$ is the effective multiplication constant for an infinite medium, accounting for virions produced per absorption without leakage loss, and $S_0(\vec{r},t)$ the virion free source.
\noindent We assume separability of the virion density into spatial and temporal components:
\begin{equation}
\label{sep}
     n_V(\vec{r},t)=f(\vec{r})n_V(t).
\end{equation}
Substituting into the diffusion equation, the spatial component satisfies the Helmholtz equation:
\begin{equation}
  \label{hel}
     \nabla^2f(\vec{r})+ B^2f(\vec{r})=0.
  \end{equation}
Where $B^2$ is the buckling factor, accounting for leakage losses. This leads to the virion kinetic equation:
\begin{equation}
  \label{ded}
     \frac{dn_V(t)}{dt}=\frac{k_\infty-(1+L^{2}B^{2})}{\ell_\infty}n_V(t)+q(t),
  \end{equation}
which corresponds to the point reactor model, applicable under uniform virion density assumptions in a homogeneous medium.

\vspace{10pt}
\noindent\textbf{Definition and Expressions of Reactivity $\rho$ and Generation Time $\ell$}

\noindent The reactivity $\rho$ is defined as the ratio of the net growth in number of virions to the number of virions produced, and the generation time $\ell$
is the average time between successive virion cycles (generations), which is inversely proportional to the virion production rate.
\noindent The two key kinetic parameters are expressed as:
\begin{equation}
  \label{lr}
      \rho=\frac{k_{eff}-1}{k_{eff}}, \ \ \ell=\frac{\ell_{0}}{k_{eff}}.
  \end{equation}
Where $k_{eff}=k_\infty/(1+L^{2}B^{2})$ is the effective multiplication constant, $\ell_0=\ell_\infty/(1+L^{2}B^{2})$ is the mean virion lifetime, $L^2=D/\Sigma_a$ defines the diffusion length $L$, and $1/(1+L^2B^2)$  represents the fraction of virions which escape leakage.
These can also be expressed in terms of production and loss rates:
\begin{equation}
 \label{rl2}
      \rho=\frac{\nu\Sigma_r-(\Sigma_a+DB^2)}{\nu\Sigma_r}, \ \  \ell=\frac{1}{\upsilon\nu\Sigma_r} .
\end{equation}
where $\nu$ is the average number of virions produced per replication event, and $\Sigma_r=\sigma_{r}N_H$ is
the macroscopic replication cross section. $\upsilon\nu\Sigma_r$ and $\upsilon{D}B^2$ denote the virion production rate and virions loss rate due to leakage, respectively.
By substituting Eq.(\ref{lr}), Eq. (\ref{ded}) becomes:
\begin{equation}
  \label{def}
     \frac{dn_V(t)}{dt}=\eta n_V(t)+q(t), \ \eta=\frac{\rho}{\ell}.
  \end{equation}
Where $\eta$ is the viral exponent.

\vspace{10pt}
\noindent\textbf{Virion Interaction Cross Sections with Host Cells}

\noindent Drawing an analogy to quantum scattering, we introduce microscopic cross sections to quantify the probability of interaction between virions and host cells, antibodies, or drug molecules. Key cross sections include:
Replication cross section $\sigma_r$, probability of successful viral entry, replication, and release;
Absorption cross section $\sigma_a$, probability of virion absorption by host cells; Capture cross section $\sigma_c$, probability of virion capture without replication. These satisfy $\sigma_a$=$\sigma_r$+$\sigma_c$.  In the initial model, we assume $\sigma_c\approx0$, simplifying to $\sigma_r=\sigma_a$.

\vspace{10pt}
\noindent\textbf{ Host Cell Burnup Equation}

\noindent Assuming host cells (H-particles) are consumed during viral replication or capture, and that this consumption does not significantly affect virion flux, we derive the burnup equation:
\begin{equation}
  \label{burne1}
     \frac{dN_{H}}{dt}=-\sigma_{a}\phi N_{H}.
\end{equation}
Where $N_H$ is the host cell number density, $\phi=\upsilon n_V$ is the virion flux.
The solution is:
\begin{equation}
  \label{burn}
  N_{H}(t)=N^{0}_{H}exp(-\sigma_{a}\upsilon\int_{0}^{t} n_V(t)dt).
\end{equation}
Where the $N^0_H$ is the initial number density, $ \upsilon$ is the velocity of virions.

\vspace{10pt}
\noindent\textbf{Chaotic Equations of Viral Production and Transmission}

\noindent The viral load, $n_V(t)$, grows exponentially at a rate characterized by a positive initial viral exponent, $n_V(t)=n_0 exp(\eta_0t)$, and can be interpreted as the chaotic divergence quantity. Within the framework of chaos theory \upcite{Ott1993,Bonasera1995}, this defines microscopic viral production chaotic system modeled as:
\begin{equation}
n_{V}(t)=\frac{n_0 n_\infty}{n_0+n_\infty exp(-\lambda_V(t))},
 \label{nv}
\end{equation}
where $\lambda_V=\eta_0=\rho_0/\ell$ is the Lyapunov Exponent governing the chaotic growth, $n_0$ is initial perturbation and $n_\infty$ is final saturation.
Similarly, the exponential growth of infected individuals is described by
$dN_{pos}/dt=\lambda N_{pos}$, with $\lambda=\chi\eta_0$,
where $N_{pos}$ denotes the number of positive cases. By interpreting $N_{pos}$  as the separation between two nearby trajectories in phase space, we can frame the population-level transmission dynamics as chaotic, characterized by the Lyapunov exponent
$LE=\lambda$.
\begin{equation}
 N_{pos}(t)=\frac{N_0 N_\infty}{N_0+N_\infty exp(-\lambda{t})},
 \label{npos}
\end{equation}
which captures the saturation dynamics of an outbreak.
We postulate a power-law relationship coupling the two systems, $N_{pos}(t)\propto{n_V(t)}^\chi$, which leads to the scaling law,
$\lambda=\chi\lambda_V$, where $\chi$ is a dimensionless scaling exponent determined from empirical data. Here, $\chi$ is not a pure fitting parameter but carries biological meaning: it can be a composite reflection of the "effective dimension" of numerous factors such as individual contact rates, viral shedding, or environmental survival.
By analyzing the transmission process within the framework of complex networks, power-law growth in population transmission is not an assumption but an inevitable consequence. Empirical studies show that in epidemics like COVID-19, the distribution of secondary cases caused by a single infected individual follows a power-law distribution \upcite{Gursakal2020}. The power-law hypothesis does not stem from the assumption itself, rather, it can emerge naturally from more fundamental mechanisms of individual heterogeneity and network spreading dynamics. Abstracting the population as a contact network, where the network itself has a power-law degree distribution, spreading dynamics on such networks have been rigorously proven to exhibit either exponential or power-law initial growth depending on the network structure\upcite{Li2018}. The V-Dynamics scaling law $\lambda=\chi\eta_0$ is remarkably consistent with this finding, suggesting that $\chi$ is essentially a measure of network topology and transmission heterogeneity.

\vspace{10pt}
\noindent\textbf{Experimental Measurements: In Vitro Cross Section}

\noindent To enable pre-outbreak prediction, we propose in vitro measurement of virion interaction cross sections using an integrated experimental system combining:
super-resolution microscopy, microfluidic platforms, single-virion tracking, and real-time data acquisition and AI-enhanced analysis.
Measured cross sections ($\sigma_r, \sigma_a^A, \sigma_a^D$) allow computation of $\rho$ and $\ell$, enabling prediction of transmission rates and virulence based on first principles. This represents a shift from descriptive to predictive virology grounded in quantifiable biophysical constants.
Using Eq.(\ref{rl2}) and considering drug effect, the $\eta$ may be computed as,
\begin{equation}
  \label{rhorate2}
  \eta=\rho/\ell=\upsilon(\nu\sigma_r N_H-\sigma_a^{A} N_A-\sigma_a^{D}N_D-DB^2).
  \end{equation}
Where $N_A$ is antibody density, $N_D$ is drug particle density. Terms in the equation represent contributions respectively from virion production, immune response, drug efficacy, and leakage loss.

\vspace{10pt}
\noindent\textbf{Calculation of Cross Section for Virus-cell Interactions}

\noindent We present a method for computing cross sections applicable to reactions between biological particles. The microscopic cross section represents the probability of a specific interaction as a function of the incident particle's kinetic energy. From a quantum mechanical perspective, the cross section for an incident particle striking a target is given by $\pi(\lambda_{d}+R)^2$, where $R$ is the target radius and $\lambda_{d}$ is the de Broglie wavelength of the incident particle, expressed as $\lambda_{d}=h/\sqrt{2Em}$, and  the Planck's constant, $h=6.626\times10^{-27}$ erg sec.
Using a virion mass of $m=1.0\times10^{-15}g$ and a physiological temperature of $T=39^0C$, $\lambda_d$ is estimated to be $\sim1.14\times10^{-13}$ cm, which is $\sim10$ orders of magnitude smaller than the diameter of a host (H) cell. Thus, for thermal virions, the wave-like contribution to the cross section is negligible - unlike in nuclear collisions (e.g., neutron-nucleus interactions), where it may dominate.
Moreover, the cross sections for virion replication or adsorption are not accurately described by the geometric area $\pi R^2$ of the host cell. We therefore introduce an effective cross-sectional area tailored to biological collisions.
For SARS-CoV-2, cellular entry is mediated by binding of the viral spike glycoprotein to angiotensin-converting enzyme 2 (ACE2) receptors on the host cell surface. The key interface is the receptor-binding domain (RBD) of the spike protein. We define the effective cross-sectional area as the total target area presented by accessible ACE2 receptors to an approaching virion, approximated as the product of the RBD area and half the number of ACE2 receptors per cell. The RBD diameter is approximately 4.8 - 4.9 nm \upcite{Cecylia2021}, corresponding to an area of $1.85 \times10^{-13} cm^2$.
The number of ACE2 receptors per cell varies widely across individuals, ranging from fewer than ten to several hundred \upcite{Xu2020}. Adopting an estimate of 300 ACE2 receptors per cell, the resulting adsorption cross section is $\sigma_a=2.775\times10^{-11}cm^2$, roughly four orders of magnitude smaller than the geometric cross section $\pi{R}^2$.
We note that $\sigma_a$ represents the cross section for virus binding to the cell surface, not necessarily for successful internalization. While ACE2 is essential for SARS-CoV-2 entry \upcite{Hoffnann2020}, post-binding entry mechanisms and their quantitative effects on infection efficiency remain to be fully incorporated into this cross section model.



\end{document}